\documentclass[twocolumn,a4paper,superscriptaddress,floatfix,showpacs,aps]{revtex4}
\usepackage[english]{babel}
\usepackage[utf8]{inputenc}
\usepackage{amsmath}
\usepackage{graphicx,epstopdf}
\usepackage{amssymb,epsfig,color,textcase}

\newcommand{\rb}{Rb$_3$Ni$_2$(NO$_3$)$_7$}

\begin{document}

\title{Electronic structure and magnetic properties of the strong-rung spin-1 ladder compound Rb$_3$Ni$_2$(NO$_3$)$_7$} 
\author{Z. V. Pchelkina}
\email{pchelkzl@mail.ru}
\affiliation{M.N. Miheev Institute of Metal Physics UB RAS, Ekaterinburg 620137, Russia}
\affiliation{Ural Federal University, Ekaterinburg 620002, Russia}

\author{V. V. Mazurenko}
\affiliation{Ural Federal University, Ekaterinburg 620002, Russia}

\author{O. S. Volkova}
\affiliation{Ural Federal University, Ekaterinburg 620002, Russia}
\affiliation{Lomonosov Moscow State University, Moscow 119991, Russia}
\affiliation{National University of Science and Technology ``MISiS'', Moscow 119049, Russia}

\author{E. B. Deeva}
\affiliation{Lomonosov Moscow State University, Moscow 119991, Russia}

\author{I. V. Morozov}
\affiliation{Lomonosov Moscow State University, Moscow 119991, Russia}

\author{S. I. Troyanov}
\affiliation{Lomonosov Moscow State University, Moscow 119991, Russia}

\author{J. Werner}
\affiliation{Kirchhoff Institute of Physics, Heidelberg University,  Heidelberg D-69120, Germany}

\author{C. Koo}
\affiliation{Kirchhoff Institute of Physics, Heidelberg University,  Heidelberg D-69120, Germany}

\author{R. Klingeler}
\affiliation{Kirchhoff Institute of Physics, Heidelberg University, Heidelberg D-69120, Germany}
\affiliation{Centre for Advanced Materials, Heidelberg University, Heidelberg D-69120, Germany}

\author{A. N. Vasiliev}
\affiliation{Ural Federal University, Ekaterinburg 620002, Russia}
\affiliation{Lomonosov Moscow State University, Moscow 119991, Russia}
\affiliation{National University of Science and Technology ``MISiS'', Moscow 119049, Russia}

\date{\today}

\begin{abstract}
Small single crystals of Rb$_3$Ni$_2$(NO$_3$)$_7$ were obtained by crystallization from anhydrous nitric acid solution of rubidium nitrate and nickel nitrate hexahydrate. The basic elements of the crystal structure of this new compound are isolated spin-1 two-leg ladders of Ni$^{2+}$-ions connected by (NO$_3$)$^-$ groups. The experimental data show the absence of long range magnetic order at T $\geq 2$~K. LDA+U calculations and the detailed analysis of the experimental data, i.e. of the magnetic susceptibility, the specific heat in magnetic fields up to 9~T, the magnetization, and of the high-frequency electron spin resonance data, enable quantitative estimates of the relevant parameters of the $S=1$ ladders in \rb . The rung-coupling $J_1 = 10.5$~K, the leg-coupling $J_2=1.6$~K, and the uniaxial anisotropy $|A| = 179$~GHz are obtained. The scenario of spin liquid quantum ground state is further corroborated by quantum Monte Carlo simulations of the magnetic susceptibility.

\end{abstract}
\maketitle

\section{Introduction}

The experimental and theoretical search for materials that are physical realizations of quantum spin models on low-dimensional lattices is a main focus of current condensed matter physics. Depending on the lattice geometry, such model materials with spin-$\frac{1}{2}$ demonstrate numerous nontrivial magnetic phenomena, like Bose-Einstein condensation of the magnons~\cite{BaCuSiO}, plateaus of the magnetization~\cite{Shastry-Satherland}, the resonating valence bond ground state (triangular lattice), skyrmions~\cite{Rosales} and others. In the recent years, prototypical low-dimensional magnetic systems with spin larger than one-half have attracted attention as experimental and theoretical studies on systems with S$\geq$1 open a way to probe the quantum states of model systems which differ from the extreme case of the spin-$\frac{1}{2}$ analogs. 

It was an unexpected theoretical discovery that coupling of two spin-$\frac{1}{2}$ Heisenberg quasi-ordered chains (with infinite correlation lengths) into spin-$\frac{1}{2}$ ladder leads to finite-range correlations and an excitation gap (for a review see Ref.~\cite{Dagotto1996}). The spin-1 single chain displays quite different properties as compared to the spin-$\frac{1}{2}$ chain, too. Its coupled ladder version, i.e. the spin-1 $N$-leg ladder has been the object of active theoretical research in the last years~\cite{Allen2000, ramos2014, Pollmann2012, Schliemann2012, Wierschem2014}. Like in the chains, the integer and non-integer spin ladders exhibit strongly different properties: using the density matrix renormalization group (DMRG) method it was shown that for semi-integer spin ladders the spin excitations are gapless for odd legs and gapped for even leg numbers. For integer spin ladders the spin gap is nonzero for both odd and even number of legs~\cite{ramos2014}. In particular, the even-leg spin-1 ladder has been found to host a symmetry-protected topological ground state~\cite{Pollmann2012}. The presence of anisotropy has been suggested to yield a nontrivial entanglement spectrum even in the unperturbed ground state~\cite{Schliemann2012}. 

Recently, the phase diagram of spin-1 weakly coupled antiferromagnetic (AFM) chains with single-ion anisotropy was obtained using the quantum Monte Carlo (QMC) method~\cite{Wierschem2014}. Experimentally, several Haldane gap compounds with uniaxial anisotropy like PbNi$_2$V$_2$O$_8$ and SrNi$_2$V$_2$O$_8$ perfectly fit to the phase boundary of this diagram. The $S=1$ ladder structure with $J_{\rm leg}\neq J_{\rm rung}$ (here $J_{\rm leg}$ and $J_{rung}$ are the magnetic exchange parameters along the legs and the rungs of the ladder, respectively) also offers the possibility to investigate  experimentally the crossover from coupled antiferromagnetic dimers to the 2-leg Haldane-like ladder. Both models are characterized by a spin-gap in the magnetic excitation spectrum and there is a crossover of the gap and the respective spin-liquid states. 

In this paper we report on the structural, electronic and magnetic properties of the first synthesized rubidium-nickel nitrate  Rb$_3$Ni$_2$(NO$_3$)$_7$. By means of thermodynamic and resonance measurements and first-principles numerical simulations we show that this compound is the physical realization of the strong-rung spin-1 ladder model. The magnetic susceptibility reveals a maximum at about 11 K, which corresponds to singlet-triplet excitations. High-frequency electron spin resonance data clearly prove significant zero-field splitting which is associated with uniaxial magnetic anisotropy of $A=-8.6$~K. The exchange interactions along the legs, the rungs, and between the ladders are numerically obtained along with their microscopic explanation within Green's function method. The experimental data obtained on a multitude of randomly oriented single crystals are compared with simulations by means of the Heisenberg model for ladders and independent dimers by means of exact diagonalization and quantum Monte Carlo methods, respectively. The results support the strong-rung spin-1 ladder scenario with $J_1 = 10.5$~K and $J_2=1.6$~K. 

\section{Experiment}

\subsection{Synthesis} The crystalline samples of Rb$_3$Ni$_2$(NO$_3$)$_7$ were synthesized from solution of rubidium nitrate RbNO$_3$ and nickel nitrate hexahydrate, Ni(NO$_3$)$_2\cdot$6H$_2$O, in anhydrous nitric acid. The solution was placed into an evacuated desiccator above the phosphorus anhydrate P$_2$O$_5$ and crystallization continued for few weeks till complete removal of the liquid phase. The molar ratio of RbNO$_3$:Ni(NO$_3$)$_2\cdot$6H$_2$O=2:1 was used, since for the stoichiometric composition of the initial mixture (3:2) magnetic admixture of Ni(NO$_3$)$_2$ was formed together with the main product. The details of the synthesis method of ammonium nitratometallates similar to that used to prepare the title compound are given in Ref.~\onlinecite{Morozov2003}.

The green crystals of Rb$_3$Ni$_2$(NO$_3$)$_7$ were mechanically separated from the colorless crystals of the rubidium nitrate present in the precipitate. It should be noted that Rb$_3$Ni$_2$(NO$_3$)$_7$ is highly hygroscopic and in air it gradually decomposes into RbNO$_3$ and nickel nitrate hydrates Ni(NO$_3$)$_2\cdot n$H$_2$O (\textit{n}=2, 4, 6). Therefore, the obtained product was stored under argon in sealed ampules and manipulations for the samples preparation for various studies were performed in a glove box under dry nitrogen or argon atmosphere. During the measurements, special efforts were taken to minimize or avoid the exposure time of the sample to air.

\subsection{Crystal structure} 

Single crystal X-ray structure determination reveals Rb$_3$Ni$_2$(NO$_3$)$_7$ to crystallize in the orthorhombic space group \textit{Pnma} (No. 62) with the lattice parameters \textit{a}=8.986(1), \textit{b}=28.063(3), \textit{c}=7.269(1) \AA, at 200 K. Refinement of 155 parameters gave a goodness-of-fit of 0.970, $R_1$=0.0315, and $wR_2$=0.0616 on all data. Good agreement of the XRD pattern of powder, prepared from the sample with the theoretical diffractogram of Rb$_3$Ni$_2$(NO$_3$)$_7$, calculated according to the crystal structure data, indicates that Rb$_3$Ni$_2$(NO$_3$)$_7$ is the only Ni-containing phase in the obtained sample. The crystal structure parameters and information on the data collection and the structure refinement are given in the \textit{Supplemental Materials} at [URL will be inserted by publisher].

\begin{figure}[!htb]
\resizebox{8cm}{!}{\includegraphics{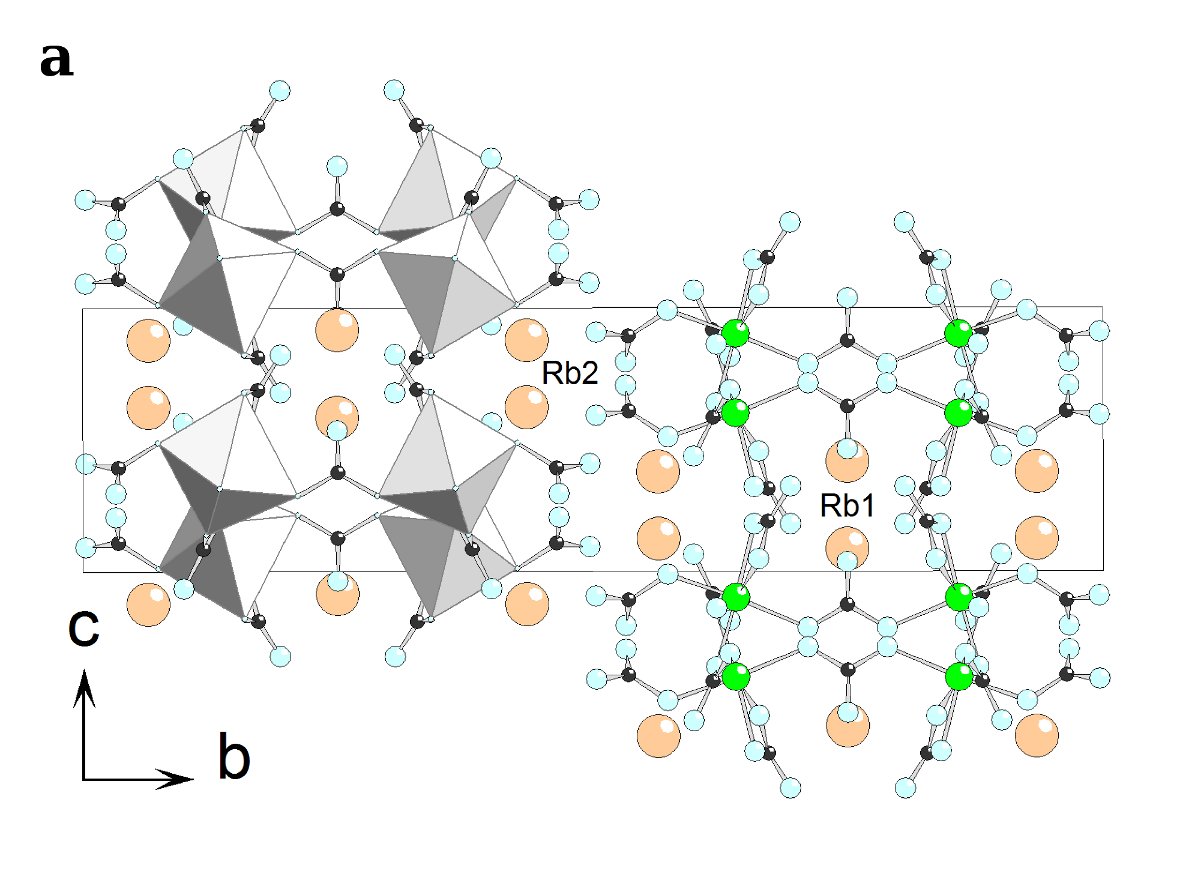}}
\resizebox{8cm}{!}{\includegraphics{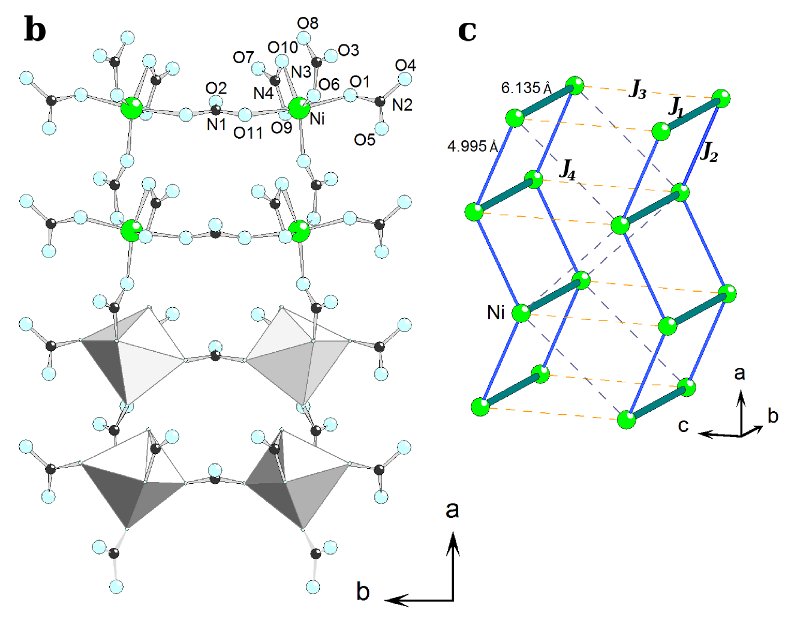}}
\caption{\label{str} (a) Projection of the Rb$_3$Ni$_2$(NO$_3$)$_7$ crystal structure along the {\textit a}-direction. (b) The NiO$_6$ octahedra are connected via NO$_3$-groups forming zigzag chains along $a$. Ni, O, N and Rb atoms are shown as green, cyan, black and orange spheres, respectively. (c) $J_1$ and $J_2$ denote exchange interactions along the rungs and the legs of the ladders, $J_3$ and $J_4$ indicate exchange interactions between the ladders.}
\end{figure}

The rubidium-nickel nitrate is isostructural to the previously synthesized (NH$_4$)$_3$Ni$_2$(NO$_3$)$_7$ compound~\cite{Morozov2003}. The crystal structure of Rb$_3$Ni$_2$(NO$_3$)$_7$ consists of zigzag [Ni$_2$(NO$_3$)$_7$]$_n^{3n-}$ ribbons with Rb$^+$ ions occupying voids between them. These ribbons have a ladder-like topology, as shown in Fig.~\ref{str}. 
The Ni$^{2+}$ ions are surrounded by distorted octahedral polyhedra formed by six oxygen atoms belonging to two terminal (mono- and bidentate) and three bridging nitrate groups. The N(1)O$_3$-group is located on a mirror plane and connects two nickel atoms by the anti-anti-type forming a rung of the ladder. Along the legs of the ladder, Ni atoms are bonded by means of syn-anti-type N(3)O$_3$-bridges resulting in a Ni$\cdots$Ni distance of 4.995~\AA~ which is much shorter than the  Ni$\cdots$Ni distance of 6.135 \AA  ~along the rungs. Selected interatomic distances and bond angles in the crystal structure of Rb$_3$Ni$_2$(NO$_3$)$_7$ are given in Tab. II and III of \textit{Supplemental Materials} at [URL will be inserted by publisher]. 
  
The distances between nickel and nearest oxygen atoms within the NiO$_{6}$-octahedra vary from 2.04 to 2.15~\AA.  In the case of the transition metal oxides, the magnetic couplings are quite sensitive to the angle of the metal-oxygen-metal bond. In Rb$_3$Ni$_2$(NO$_3$)$_7$, the NiO$_6$ octahedra are linked through NO$_3$-groups. The Ni-N-Ni angles
along the {\it a}- and the {\it b}-axis are 119$^\circ$ and 172$^\circ$, respectively.  It is illustrative to compare these values to NiO (F\emph{m-3m} structure). In NiO, the Ni-O distance is 2.1 \AA~and the Ni-O-Ni angle is 180$^\circ$ which results in a strong antiferromagnetic superexchange interaction of 221~K (19~meV)~\cite{Hutchings1972}. Taking into account the strong distortion of the NiO$_6$ octahedra in  Rb$_3$Ni$_2$(NO$_3$)$_7$ and the longer distances between neighboring octahedra coupled by nitrate groups, one could expect that the magnetic exchange interaction in this system should be much weaker than in NiO.      

\subsection{Thermodynamic properties} 

The temperature dependency of the magnetic susceptibility of Rb$_3$Ni$_2$(NO$_3$)$_7$ was measured at $B$=0.1~T by means of a Magnetic Properties Measurement System (MPMS XL-5, from Quantum Design) while the field dependence of magnetization at T=4.2~K in magnetic fields up to 15~T was studied by means of a home-built vibrating sample magnetometer (VSM)~\cite{vsm}. The specific heat was measured at various magnetic fields up to 9 T by means of a Physical Properties Measurement Systems (PPMS, from Quantum Design). 
\begin{figure}[!htb]
\resizebox{8cm}{!}{\includegraphics[angle=0]{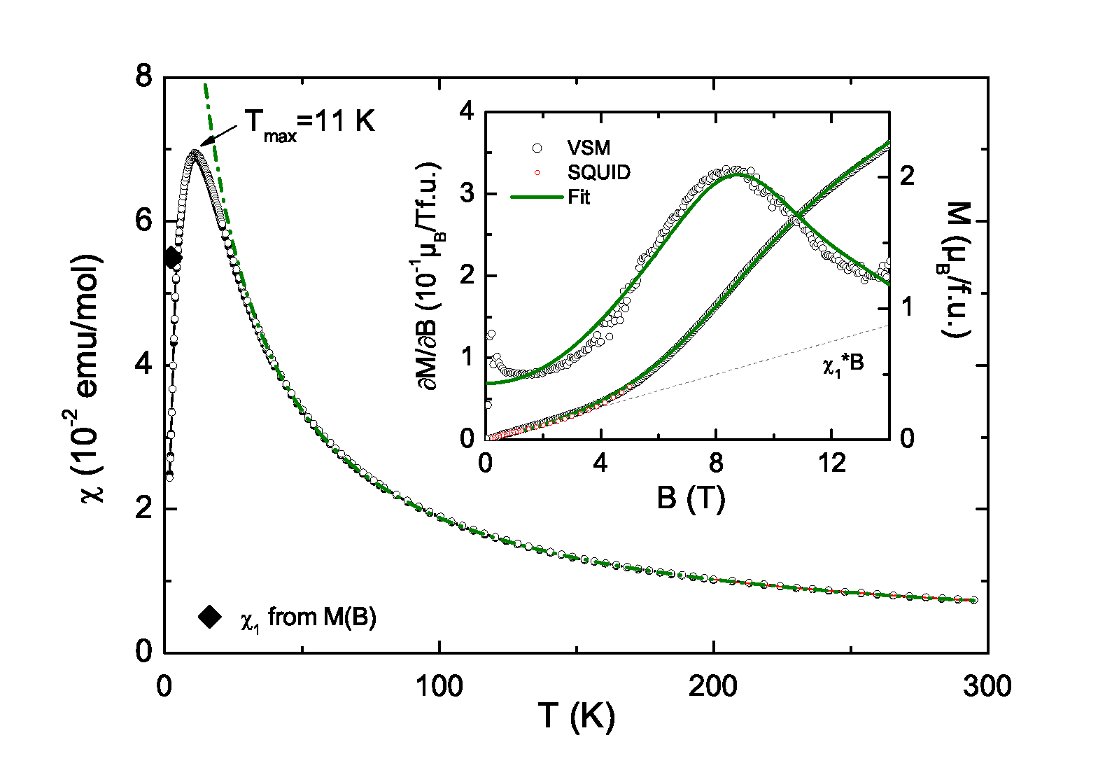}}
\caption{\label{susceptibility} 
Temperature dependence of the magnetic susceptibility of Rb$_3$Ni$_2$(NO$_3$)$_7$, at $B$=0.1~T, taken in the field-cooled regime. The dashed-dotted line shows a fit in accordance with the Curie-Weiss law. Field dependence of the magnetization, at $T$=4.2~K, and its derivative are shown in the insert. Green lines are fits to the data (see the text). The dotted line illustrates the low field susceptibility $\chi_1$.}
\end{figure}

The temperature dependence of the magnetic susceptibility $\chi$=$M/B$ of Rb$_3$Ni$_2$(NO$_3$)$_7$, $B$=0.1~T, is presented in Fig.~\ref{susceptibility}. In the whole temperature range under study, there is no difference of the measurements obtained in the field-cooled and the zero-field cooled regime which evidences the absence of any considerable ferromagnetic impurities. The low-temperature behavior implies only a few quasi-free defect spins. The analysis of the field dependence of the magnetization at small $B$ (not shown) indeed confirms a very small number of quasi-free defect spins $S=1$ of about 0.1~\%. In the temperature range 50-300~K, the $\chi$(T) dependence shown in Fig.~\ref{susceptibility} can be described by a Curie-Weiss-like behavior, $\chi$(T)=$\chi_0$+$C$/(T-$\Theta$), with a temperature independent term $\chi_0$=6.6$\cdot$10$^{-4}$~emu/mol, the Curie constant $C$=2.02~K emu/mol, and the Weiss temperature $\Theta$=-10.9~K. At lower temperatures, $\chi$(T) deviates from the Curie-Weiss behavior, passes through a maximum at $T_{\rm max}$=11~K, and subsequently drops more than twice at T$<$T$_{max}$. From the Curie constant, the effective magnetic moment $\mu_{\rm eff} = 4.0(2) \mu_B/f.u.$ is extracted, which for $S=1$ is associated with the $g$-factor $g$=2.01(5). The negative value of the Weiss temperature indicates the predominance of antiferromagnetic exchange interactions at elevated temperatures. 

The main magnetic substructures of {Rb$_3$Ni$_2$(NO$_3$)$_7$} are the two-leg ladders based on Ni$^{2+}$ ions (cf. Fig.\ref{str}c). Depending on the ratio of exchange interaction parameters on the rungs and the legs, this structure's extremities are isolated dimers~\cite{Carlin} or uniform chains~\cite{Law}. The maximum present in $\chi$(T) can be inherent to each of these cases. Indeed, the experimental data are roughly described in terms of either of these extremal cases by means of analytical expressions for dimers and uniform chains. Fitting the data  by means of both models yields the main exchange interaction parameters  $J_{\rm dim}$=11~K (pure dimer model) or $J_{\rm chain}$=8.4~K (pure chain model). As will be shown in section~\ref{comp}, the generalized ladder model yields a significantly better description of the data.

\begin{figure}[!ht]
\resizebox{8cm}{!}{\includegraphics[angle=0]{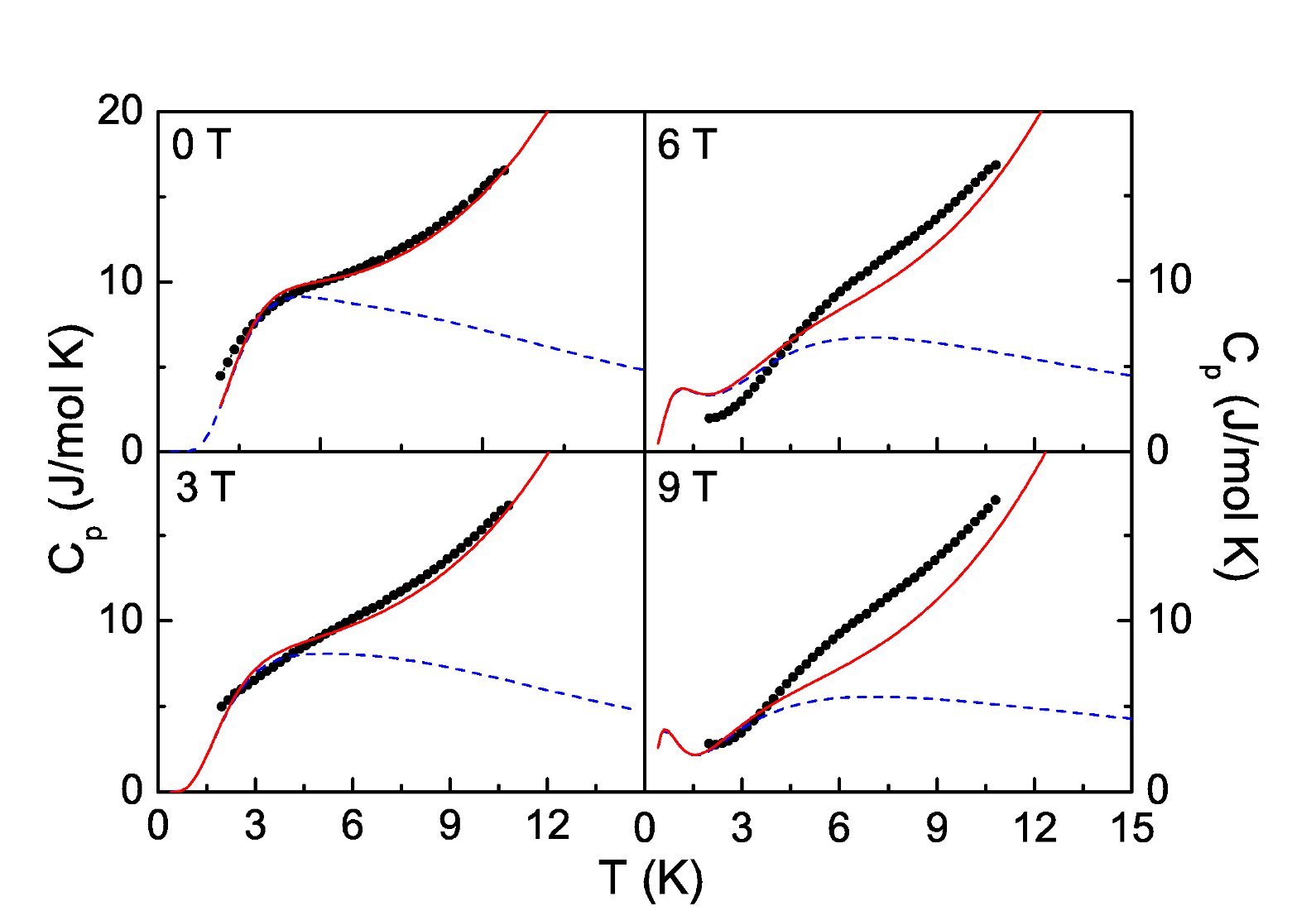}}
\caption{\label{specific_heat} The temperature dependencies of the specific heat of Rb$_3$Ni$_2$(NO$_3$)$_7$ measured at various magnetic fields. The dotted lines were calculated in the model of $S$=1 dimers. Solid lines represent the sum of the $S$=1 dimer contribution and a T$^3$ term.}
\end{figure}
The magnetization curve $M$($B$) taken in quasistatic magnetic field and its derivative $\partial M(B)/\partial B$ are shown in the inset of Fig.~\ref{susceptibility}. The $M$($B$) curve bends at about $B_C$=10 T seen as a maximum in the $\partial M/\partial B$ dependence. At $B\!>\!B_C$, the magnetization approaches $\sim$ 2$ \mu_B$ which is about half of the expected saturation magnetization $M_{\rm sat}$=$ng$S$\mu_B\!\approx$4 $\mu_B$/f.u.. In a dimer-like model, left-bending of the magnetization signals field induced changes of the lowest spin energy state(s) providing information on the energy difference of the singlet and the lowest triplet  state. We note that a spin gap in isolated $S$=1 chains as well as in $S$=1 spin ladders implies a corresponding anomaly in $M$ vs. $B$, too, as, e.g., seen in NENP~\cite{NENP}. Quantitatively, the $S$=1 uniform chain model presumes the magnetic field necessary to overcome energy gap $\Delta$ ($\Delta_{chain}$=0.41$J_{chain}$) to be equal to $B$=2.56~T which is significantly smaller than the experimentally found value.

\begin{figure}[!ht]
\resizebox{8cm}{!}{\includegraphics[angle=0]{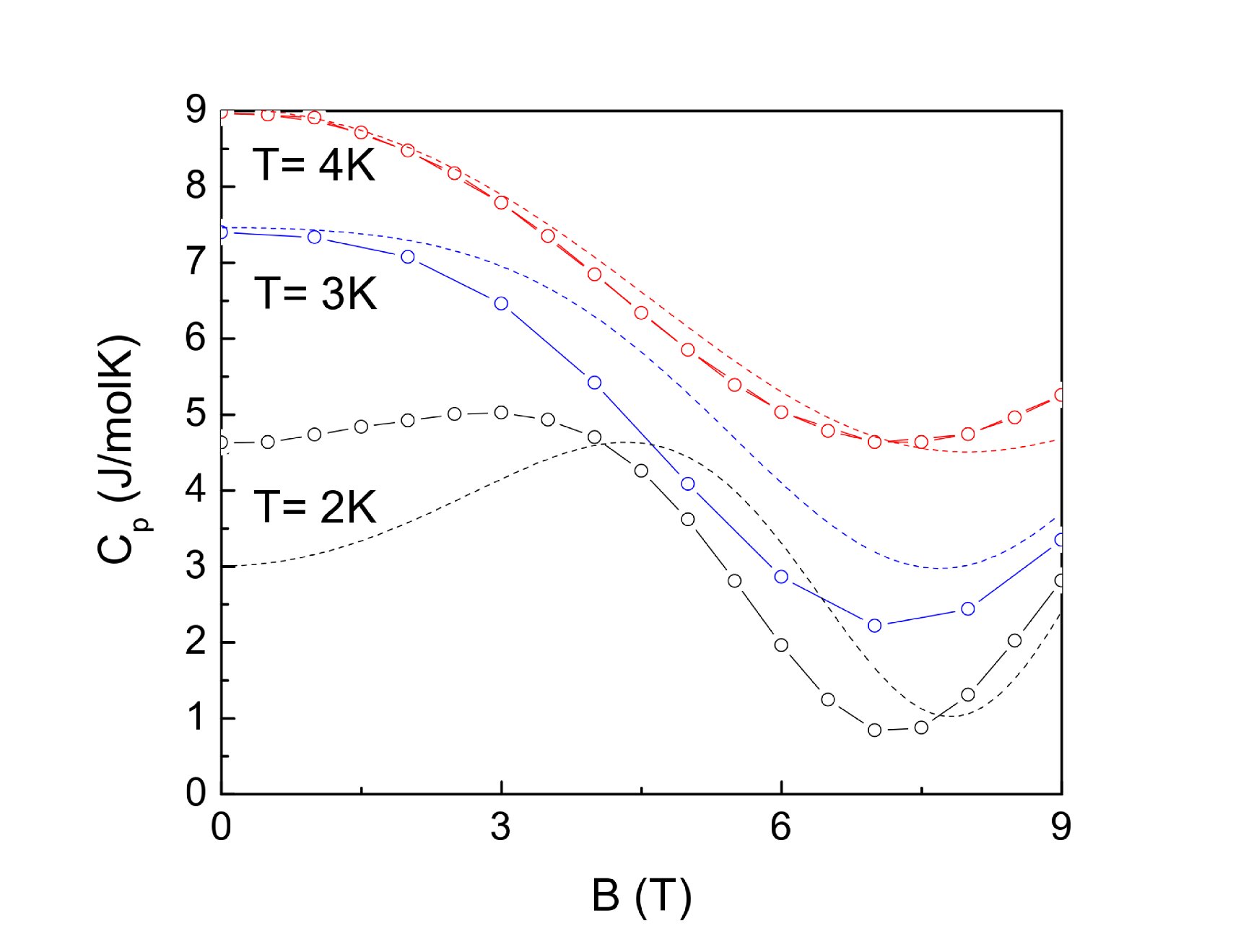}}
\caption{\label{specific_heat_t} The field dependencies of the specific heat of Rb$_3$Ni$_2$(NO$_3$)$_7$ at several constant temperatures. The circles are the experimental data, dotted lines show the expected behavior in the dimer model (see the text).}\end{figure}

The temperature dependencies of the specific heat $C_p$(T) of Rb$_3$Ni$_2$(NO$_3$)$_7$ taken at various magnetic fields up to 9 T are shown in Fig.~\ref{specific_heat}. The $C_p$(T) curve at $B$=0 clearly shows a low temperature Schottky anomaly. This anomaly can be attributed to the presence of energetically separated $S$=0, 1, 2 levels in the energy spectrum of Ni$^{2+}$ dimers which become thermally populated upon heating (cf. Fig.~\ref{epr2}). Upon application of external magnetic fields, the ground state energies are shifted by the Zeeman effect which yields, e.g., the abovementioned changes of the ground states showing up in the anomaly in $M$($B$). Indeed, the Schottky anomaly clearly changes upon application of external magnetic fields. This effect is most clear if the magnetic field dependence of $C_{\rm p}$ at constant temperature is considered (see Fig.~\ref{specific_heat_t}). Here, the specific heat is strongly suppressed at about $B$=$B_{\rm C}$, i.e. it signals the crossing of the ground state spin levels. 

A quantitative estimate of the magnetic contribution to the specific heat uses the associated partition function, the mean energy and the mean squared energy for $\Delta_{\rm dimer}$=11~K and $g$=2.01 obtained from the magnetic measurements. The resulting curves are shown by dotted lines in Fig.~\ref{specific_heat}. The lattice contribution to the specific heat was considered by $C_{\rm p}^{\rm lattice}$=$\beta$T$^3$, with $\beta$=$ 1943.7\cdot s/\Theta_{\rm D}^3$. Here, $\Theta_{\rm D}$ is the Debye temperature and $s$=33 the number of atoms per Rb$_3$Ni$_2$(NO$_3$)$_7$ formula unit. The specific heat of Rb$_3$Ni$_2$(NO$_3$)$_7$ was measured up to 200~K (data not shown) where we obtained $C_{\rm p}\approx 610$~J/(mol$\cdot$K) which is about 75\% of the expected Dulong-Petit limit 3$R \cdot$s=823~J/(mol$\cdot$K). Thus, our experimental data imply a lower limit of $\Theta_D\!>$200~K for the Debye temperature and accordingly an upper limit for $\beta$ which is smaller than 0.008 J/(mol$\cdot$K$^4$). The resulting dimer and lattice contributions to the specific heat shown by the solid lines in Fig.~\ref{specific_heat} describe the experimental data quite well in the temperature regime T$>$T$_{\rm max}$. In contrast, there are notable discrepancies at low temperatures T$<$T$_{max}$, where the lattice contribution is negligibly small. Tentatively, this indicates a more complex scenario than the simple non-interacting $S$=1 dimer model. 

\subsection{High-frequency ESR spectroscopy}

High-frequency/high field Electron Spin Resonance (HF-ESR) measurements were carried out using a phase-sensitive millimeter-wave vector network analyzer from AB Millimetr\'e covering the frequency range from 30 to 1000 GHz and in magnetic fields up to 18~T~\cite{Comba2015}. For the experiments, a fixed powder sample in an airtight glass vessel was placed in the sample space of the cylindrical waveguide. 

\begin{figure}
\includegraphics[width=1.0\columnwidth,clip] {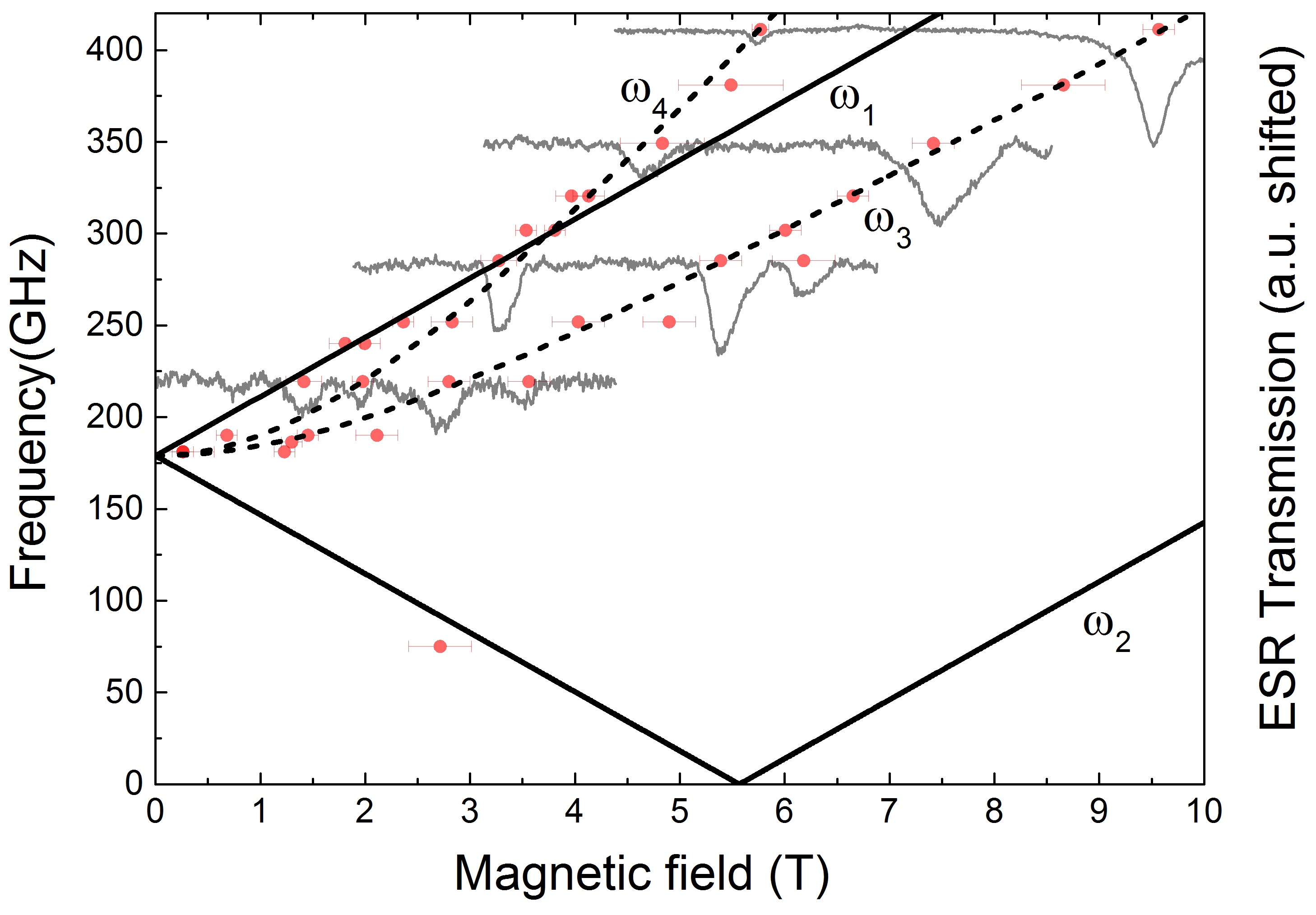}
\caption{\label{epr1} HF-ESR transmission spectra (shown in gray), at $T=3$~K, with corresponding resonance positions marked by red dots in the frequency vs. magnetic resonance field diagram. The lines are the simulated resonance branches obtained by solving the spin Hamiltonian (Eq.~(\ref{ESRHamil})). Solid lines correspond to $\Theta$=0$^{\circ}$ and dashed ones to $\Theta$=90$^{\circ}$ (see the text).}
\end{figure}

Typical ESR spectra of \rb\ at $T$=3~K exhibit broad and separated ESR lines as shown by some examples in Fig.~\ref{epr1}. The spectra allow collecting the respective frequency  vs. resonance field $B_{res}$ dependencies of the observed lines which are denoted by the red dots. Even without further analysis, the data imply a significant zero field splitting (ZFS) of around $\Delta \approx 180$~GHz. Our quantitative analysis applies the dimer model of Ni$^{\rm 2+}$-spins $S_1$=$S_2$=1 with uniaxial anisotropy $A_1$=$A_2$=$A$, i.e.:

\begin{equation}
	\label{ESRHamil}
	\hat{H}=J \hat{S}_1 \hat{S}_2+A (\hat{S}{_1^z}^2+\hat{S}{_2^z}^2)+g \mu_B B(\hat{S}_1+ \hat{S}_2).
\end{equation}
Here, $J$ is the intra-dimer exchange coupling, $B$ the external magnetic field, and $g$ the effective $g$-factor. Numerical evaluation of the Hamiltonian~(\ref{ESRHamil}) has been performed by means of the EasySpin toolbox for Matlab~\cite{stoll2006}. 

\begin{figure}
\includegraphics[width=1.0\columnwidth,clip] {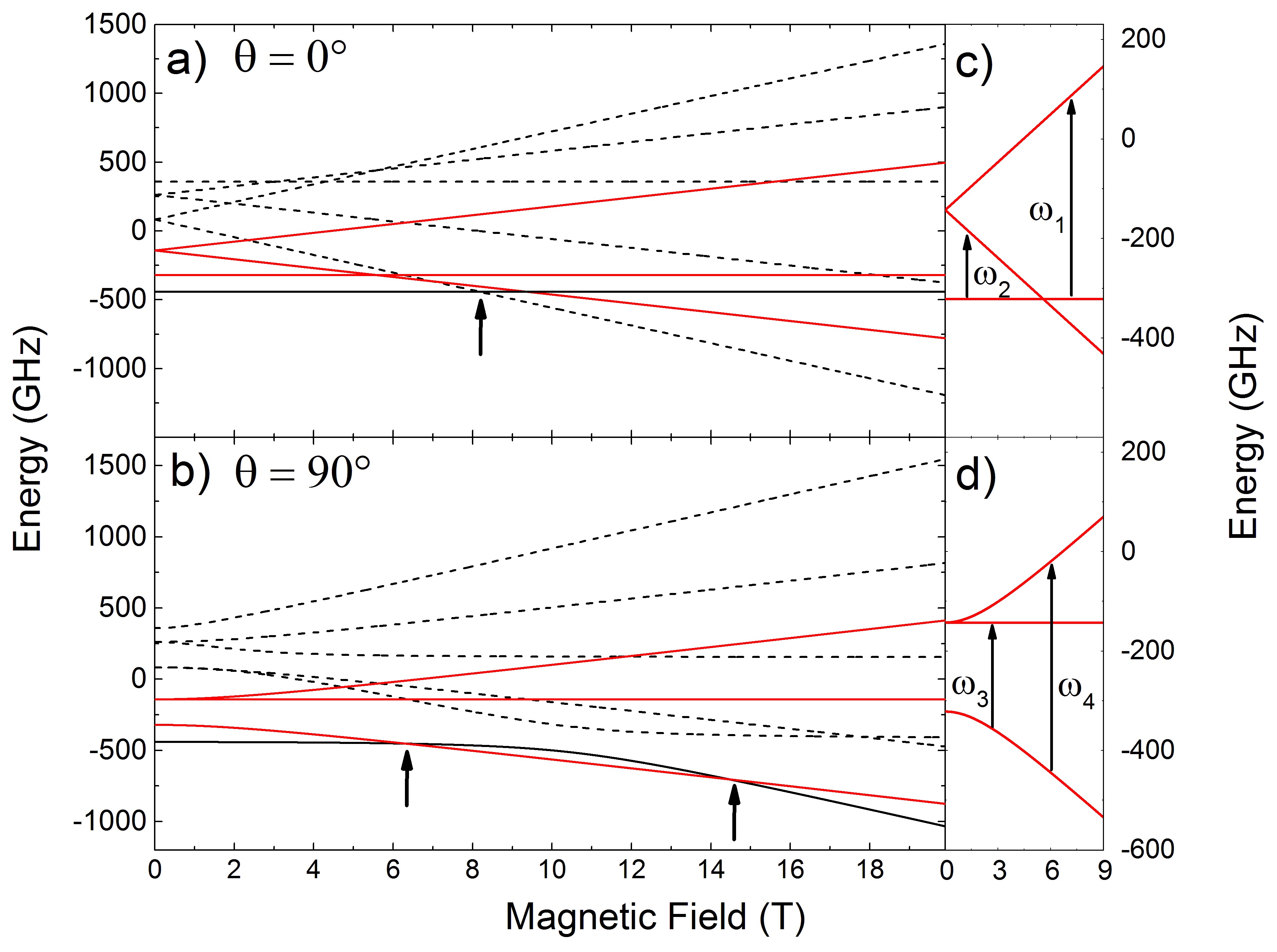}
\caption{Energy level diagrams obtained by solving Eq.~(\ref{ESRHamil}) for the single ion anisotropy axis being parallel (a) and orthogonal (b) to the external magnetic field. Black, red, and dashed lines show the singlet, triplet, and quintuplet states, black arrows indicate ground state level crossing. (c) and (d) show the energy levels of the triplet states in which the observed HF-ESR transitions associated with the resonance branches $\omega_1$ to $\omega_4$ (cf. Fig.~\ref{epr1}) occur.}\label{epr2}
\end{figure}

The measured powder sample implies that for a given magnetic field direction the angle $\Theta$ between $B$ and the local single ion anisotropy axis varies as $0^{\circ}\!\leq\!\Theta \leq 90^{\circ}$. The effect of $\Theta$ on the energy level diagram is illustrated in Fig.~\ref{epr2} which displays the situation in the extreme cases of the single ion anisotropy being parallel and orthogonal to the external magnetic field, respectively. Note, that without mixing of the spin states, the selection rules $\Delta m_S$=$\pm 1$ allow transitions only within a given multiplet, while $\Delta S$=$\pm 1$ (e.g., singlet to triplet transitions) are forbidden. At the temperature of the experiments, transitions within the $S$=2 quintet states are not observed since the intradimer exchange coupling $J$ results in a too large energy difference between the spin multiplets as compared to the thermal energy, i.e. $E_{\rm triplet}$-$E_{\rm quintet}\!\ll\!k_{\rm B} T$. Thus, in our experiments, allowed transitions are supposed within the triplet states only (see Fig.~\ref{epr2}c and d). 

The lines in Fig.~\ref{epr1} display the results of the simulations. The solid lines refer to the case $\Theta$=0$^{\circ}$ and the dashed lines to $\Theta$=90$^{\circ}$. The results show that the overall behavior is well described in terms of the dimer model Eq.~(\ref{ESRHamil}). The obtained parameters are the single ion anisotropy $|A|$=179(1)~GHz (i.e., 8.6~K) and the isotropic effective $g$-factor of $g$=2.31(5), at T=3~K. Although our data do not unambiguously allow determining the sign of $A$, the following simulations apply a uniaxial case typically observed in octahedrally coordinated Ni$^{2+}$-complexes, i.e. $A\!<$0~\cite{Mennerich2006, Kry, Krupskaya, das2011new, popova2012thermodynamic, popova2013magnetic, yamaguchi1970electron}. This assumption is corroborated by the temperature dependence of the $\omega_1$-resonance branch (see the \textit{Supplemental Materials}). The size of the anisotropy corresponds well with the recently reported value of -9~K for Li$_2$NiW$_2$O$_8$ which is supposed to exhibit spin-1 Ni$^{2+}$ chains~\cite{Ranjith2016}. A similar value of the single ion anisotropy A=-11.5~K was also reported for Na$_2$Ni$_2$(C$_2$O$_4$)$_3$(H$_2$O)$_2$ which realizes the spin-1 strong rung two-leg ladder structure, too~\cite{Mennerich2006}.

The calculated energy levels shown in Fig.~\ref{epr2} allow to assign the different branches to particular transitions. At $B$=0, all observed resonance branches within the $S$=1 multiplet are degenerated. The resonance branch $\omega_1$ is associated with the transition $m_S$=0$\rightarrow$m$_S$=1, and $\omega_2$ with $m_S$=-1$\rightarrow$m$_S$=0. For $\Theta$=90$^{\circ}$, $\omega_3$ is associated with $m_S$=-1$\rightarrow$m$_S$=0. Note, that the transition $m_S$=0$\rightarrow$m$_S$=1 ($\Theta$=90$^{\circ}$) is not observed in the experiment since the $m_S$=0 spin energy state is considerably above the $m_S$=-1 state and not populated at low temperature (see Fig.~\ref{epr2}d). In contrast, the spectra show the presence of the $\Delta m_S$=$\pm 2$ transition $m_S$=-1$\rightarrow$ m$_S$=+1 showing up in the branch $\omega_4$. Accordingly, at high magnetic field the branch $\omega_4$ has almost the double slope as compared to the dipole allowed branches $\omega_1$-$\omega_3$ (cf. Ref.~\onlinecite{Mennerich2006}). The observed finite intensity of the so-called forbidden transition implies mixing of the pure spin states $m_s$=-1, $m_s$=0, and $m_s$=+1 due to crystal field and spin-orbit effects. In contrast, resonances within higher spin multiplets are not observed by our HF-ESR measurements, which agrees to large intradimer coupling yielding the higher multiplets energetically well separated from the triplet states. 

While the free dimer model of Eq.~(\ref{ESRHamil}) describes the positions and field dependencies of the resonances sufficiently well, there are deviations with respect to the experimentally observed spectral intensities including splitting of the resonance branches at small and intermediate magnetic fields $B\!\lesssim$6~T. These discrepancies imply that a more complex scenario has to be considered to fully describe the magnetic properties of \rb , e.g., by including interdimer coupling and transversal anisotropy as well as Dzyaloshinskii-Moriya interactions (see below).

Applying Eq.~(\ref{ESRHamil}) allows describing the magnetization data $M(B)$. The results are shown in Fig.~\ref{susceptibility}. In order to imitate powder averaging realized in the sample under study, the simulations were done by calculating the magnetization from the Hamiltonian (\ref{ESRHamil}) in steps of 1$^{\circ}$ increments between $\Theta$=0$^{\circ}$ and $\Theta$=90$^{\circ}$ (here $\Theta$ is an angle between magnetic field direction and the local single ion anisotropy axis) and following averaging over the magnetizations obtained for different angles. By using the values of $A$ and $g$ from the analysis of the HF-ESR data, the fits yield $J_{\rm dimer}$=9.7(8)~K. The field dependence of the spin energy states shown in Fig.~\ref{epr2} is in correspondence with magnetization curve shown in the insert of Fig.~\ref{susceptibility}. For the two extreme angles shown in Fig.~\ref{epr2}, the arrows indicate changes of the magnetic ground state. To be specific, at $\Theta$=0$^{\circ}$, there is a field induced crossing from the $|S=0,m_S=0\rangle$ to the $|S=2,m_S=-2\rangle$ state while at $\Theta$=90$^{\circ}$, the ground state successively changes to different mixed states which evolving from the $|S=1,m_S=-1\rangle$ and $|S=2,m_S=-2\rangle$ ones, respectively. The field dependence $M$ vs. $B$ is approximately explained by means of the independent dimer model applying the experimentally determined parameters $A$ and $g$, and using an appropriate interdimer coupling $J$. As will be shown in section \ref{comp}, the magnetic exchange parameters can be further elaborated in the frame of the more realistic spin ladder model.

\section{First principles calculations}

\subsection{Electronic structure} 
The electronic structure of Rb$_3$Ni$_2$(NO$_3$)$_7$ was calculated with the TB-LMTO-ASA (Tight Binding-Linearized Muffin-Tin Orbitals-Atomic Sphere Approximation) code~\cite{tb-lmto} and the Vienna Ab initio Simulation Package (VASP)~\cite{Kresse1996}. The radii of atomic spheres in the TB-LMTO-ASA calculations were chosen as follows: R(Ni)=2.3 a.u., R(O)=1.3-1.6 a.u., R(N)=1.3-1.4, R(Rb)=4.4 a.u. 
A mesh of 32 irreducible $k$-points was used in the VASP calculations. The plane wave cut-off energy was chosen to be 400 eV. The rotationally invariant form of LSDA+U method was utilized~\cite{Liechtenstein1995}. 

\begin{figure}[!htb]
\resizebox{8cm}{!}{\includegraphics[angle=-90]{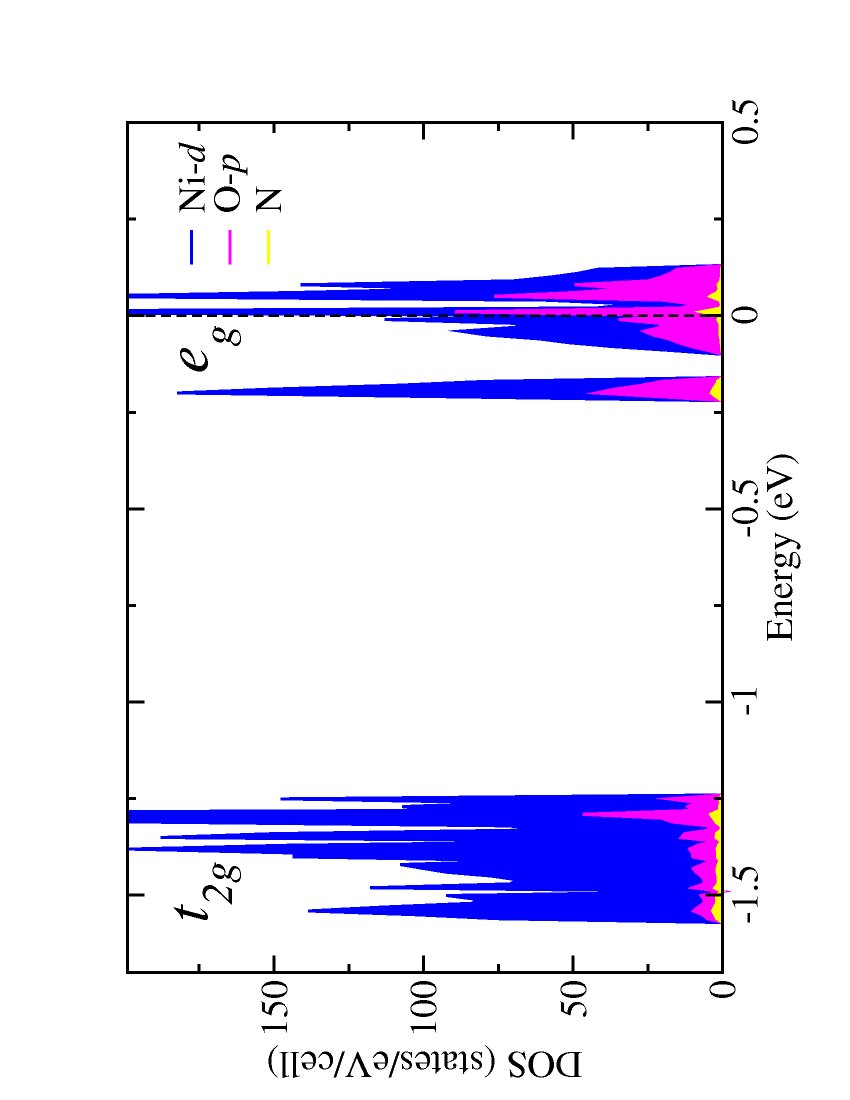}}
\caption{\label{lda_dos} The LDA Ni-$d,~$O-$p$ and N-$s,p$ partial densities of states for Rb$_3$Ni$_2$(NO$_3$)$_7$. The Fermi energy corresponds to $E$=0.}
\end{figure}

\begin{figure}[!htb]
\resizebox{8cm}{!}{\includegraphics[angle=-90]{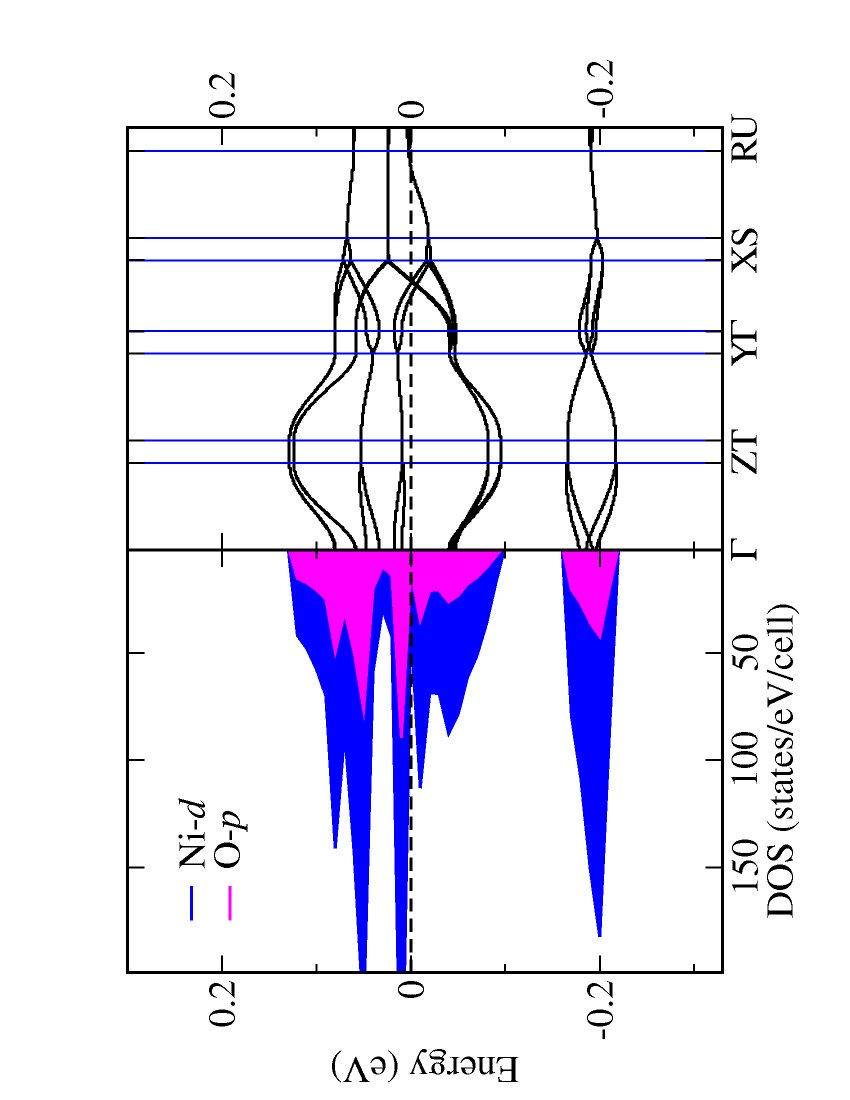}}
\caption{\label{band} LDA band structure ($e_g$ bands) around the Fermi level and corresponding partial Ni-$d, ~$O-$p$ DOS for Rb$_3$Ni$_2$(NO$_3$)$_7$. The Fermi energy corresponds to $E$=0.}
\end{figure}

The resulting partial densities of states (DOS) and band structure obtained in local density approximation (LDA) are shown in the Figures~\ref{lda_dos} and \ref{band}.  
The DOS around the Fermi level is formed by Ni-$d$ states hybridized strongly with O-2$p$ states. The $t_{2g}$~-~$e_{g}$ splitting can be estimated as 1.5~eV, which agrees with that calculated for other nickel oxides (1.1~eV for NiO, 1.7~eV for Ni(NO$_{3}$)$_2$ \cite{Volkova}). The splitting of the $e_{g}$ DOS near the Fermi level is caused by the dimerization accompanied by the formation of the bonding and antibonding states (see Fig.~\ref{band}). The width of the highest $e_{g}$ band is about 0.22~eV, hence the dimer hopping can be estimated as about  110~meV.     

To perform a quantitative analysis of the LDA results we constructed the low-energy model describing the $e_g$ bands near the Fermi level by using the projection procedure~\cite{Hproj}. The average transfer integrals were evaluated as $\overline{t}_m\!\equiv\!\overline{t}_{ij}$=$\sqrt{{\rm Tr}(\hat{t}_{ij} \hat{t}_{ij}^{T})}$, where $\hat{t}_{ij}$ is the 2$\times$2 matrix in the basis of Wannier functions constructed for the $e_g$ bands and the trace is taken over orbital indices. The notation of the hopping integrals $\overline{t}_m$ ($m$=1-4) corresponds to the notation of the main exchange paths shown in Fig.~\ref{str}c. We obtain $\overline{t}_1$=110~meV, $\overline{t}_2$=47~ meV, $\overline{t}_3$=3~meV, $\overline{t}_4$=15~meV. From these results one can already assume that Rb$_3$Ni$_2$(NO$_3$)$_7$ can be classified as a spin-1 two-leg ladder compound.

The metallic ground state of Rb$_3$Ni$_2$(NO$_3$)$_7$ obtained within LDA (see Fig.~\ref{band}) is a standard problem of this approach which is due to the underestimation of the on-site Coulomb correlations. To resolve it, we implement the LDA+U method~\cite{Anisimov1997}. The values of the on-site Coulomb repulsion parameter and the intra-atomic Hund's rule exchange interaction were chosen to be $U$=6.5-7 eV and $J_{\rm H}$=0.95~eV. As will be shown below, such a parameter choice leads to a good agreement between experiment and theory on the magnetic susceptibility. Moreover, these parameters are close to those estimated within the LDA constrained procedure for other nickel oxides~\cite{Anisimov1991, Tran2006}. 

\begin{figure}[!htb]
\resizebox{8cm}{!}{\includegraphics[angle=-90,]{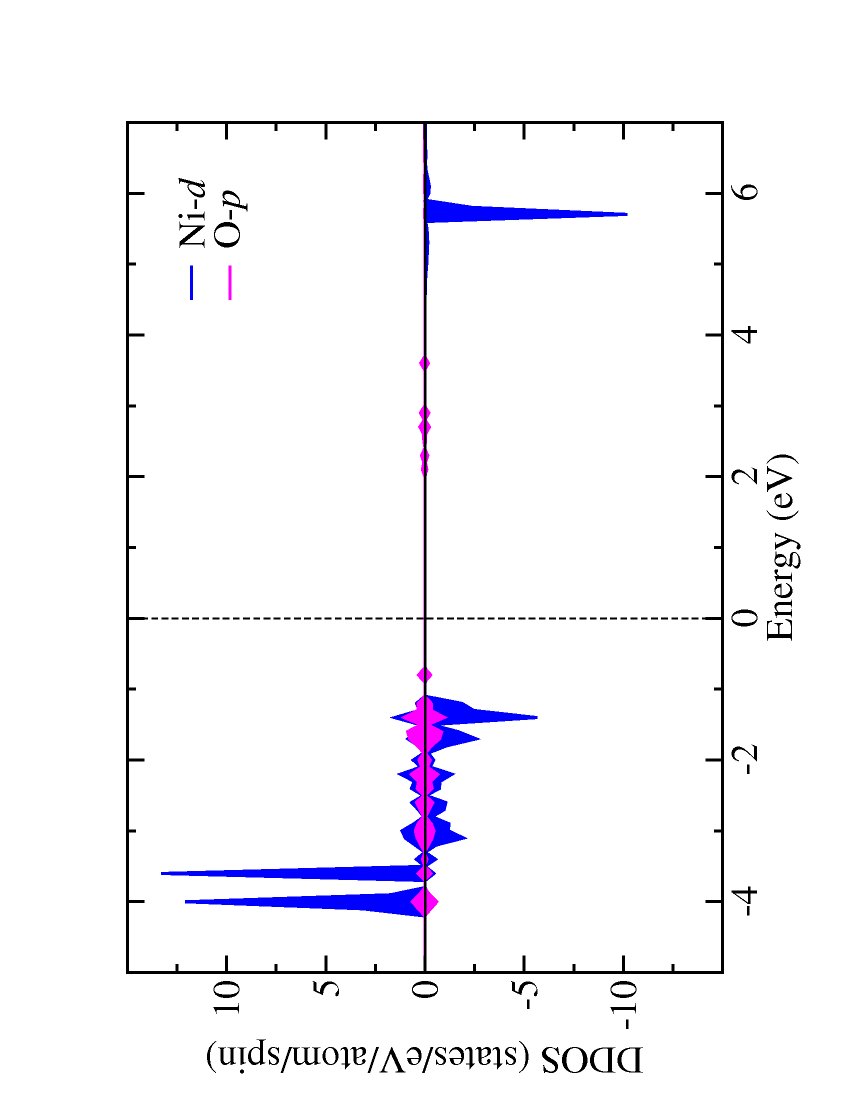}}
\caption{\label{ldau_dos} The LDA+$U$ Ni-$d$ and O-$p$ partial density of states for Rb$_3$Ni$_2$(NO$_3$)$_7$. The Fermi energy corresponds to zero.}
\end{figure}

Since the Curie-Weiss temperature of Rb$_3$Ni$_2$(NO$_3$)$_7$ is negative, antiferromagnetic arrangement of magnetic moments on Ni ions was imposed in the LDA+U calculation. We obtained an insulating ground state with the magnetic moment of 1.87-1.91~$\mu_{\rm B}$ per Ni and an energy gap of 2.06-2.11~eV for $U$=6.5-7~eV. The value of the magnetic moment is in correspondence with typical values for nickel oxides (1.9 $\mu_B$ for NiO~\cite{Cheetham83}). The partial spin-resolved Ni-$d$ DOS is shown in Fig.~\ref{ldau_dos}. Majority spin states are completely occupied while minority spin states are partially filled in accordance with the $d^{8}$ electronic configuration implied by the Ni$^{2+}$ oxidation state.  

\subsection{Magnetic interactions} 

In order to describe the magnetic properties of the Rb$_3$Ni$_2$(NO$_3$)$_7$ system we used the following Hamiltonian 
\begin{equation}
\label{Heisenberg}
H = \sum_{ij} J_{ij} \hat {\vec S}_i \hat {\vec S}_j + \sum_{ij} \vec D_{ij} [\hat {\vec S}_{i} \times \hat {\vec S}_{j}] + \sum_{i \mu \nu} \hat S^{\mu}_{i} A^{\mu \nu}_{i} \hat S^{\nu}_{i},
\end{equation}
where $J_{ij}$ is the isotropic exchange interaction, $\vec D_{ij}$ the Dzyaloshinskii-Moriya vector, and $A_{i}^{\mu \nu}$ are the elements of the single-ion anisotropy tensor, $\mu (\nu)$=$x,y,z$. The summation $\sum_{ij}$ accounts each pair once.

All parameters of the spin Hamiltonian were calculated by using the magnetic force theorem with different types of perturbations. For instance, for the isotropic exchange interactions we used the approach reported in Refs.~\onlinecite{Lichtenstein1987, Mazurenko2005}, which is based on the infinitesimal rotation of the magnetic moments from the antiferromagnetic ground state. In turn, the elements of the single-ion anisotropy tensor were obtained by using the spin-orbit coupling as a perturbation~\cite{Solovyev, Mn12}. The procedure for calculating the Dzyaloshinskii-Moriya interactions is based on the mixed type of the perturbation theory on the spin-orbit coupling and infinitesimal rotation of the magnetic moments~\cite{Mn12}.   

The main isotropic exchange pathways for Rb$_3$Ni$_2$(NO$_3$)$_7$ are shown in Fig.~\ref{str}c, where $J_1$ and $J_2$ are the couplings along the rungs and the legs of the ladder, respectively. $J_3$ and $J_4$ denote the exchange interactions between the ladders. The calculated values of the exchange integrals are $J_1$=11-9.84~K, $J_2$=1.62-1.44~K, $J_3$=0~K, and $J_4$=0.07~K for the chosen range of the on-site Coulomb repulsion parameter. All exchange interactions are antiferromagnetic. According to the calculated exchange constants, we classify Rb$_3$Ni$_2$(NO$_3$)$_7$ as a spin-1 two-leg ladder compound. The obtained ratio $\frac{J_2}{J_1}$=0.15 indicates very strong rung interaction. This ratio is about two times smaller than what was experimentally found for the spin-1/2 ladder analog (C$_5$H$_{12}$N)$_2$CuBr$_4$~\cite{CuBr4}.

\begin{figure}[!htb]
\resizebox{7cm}{!}{\includegraphics{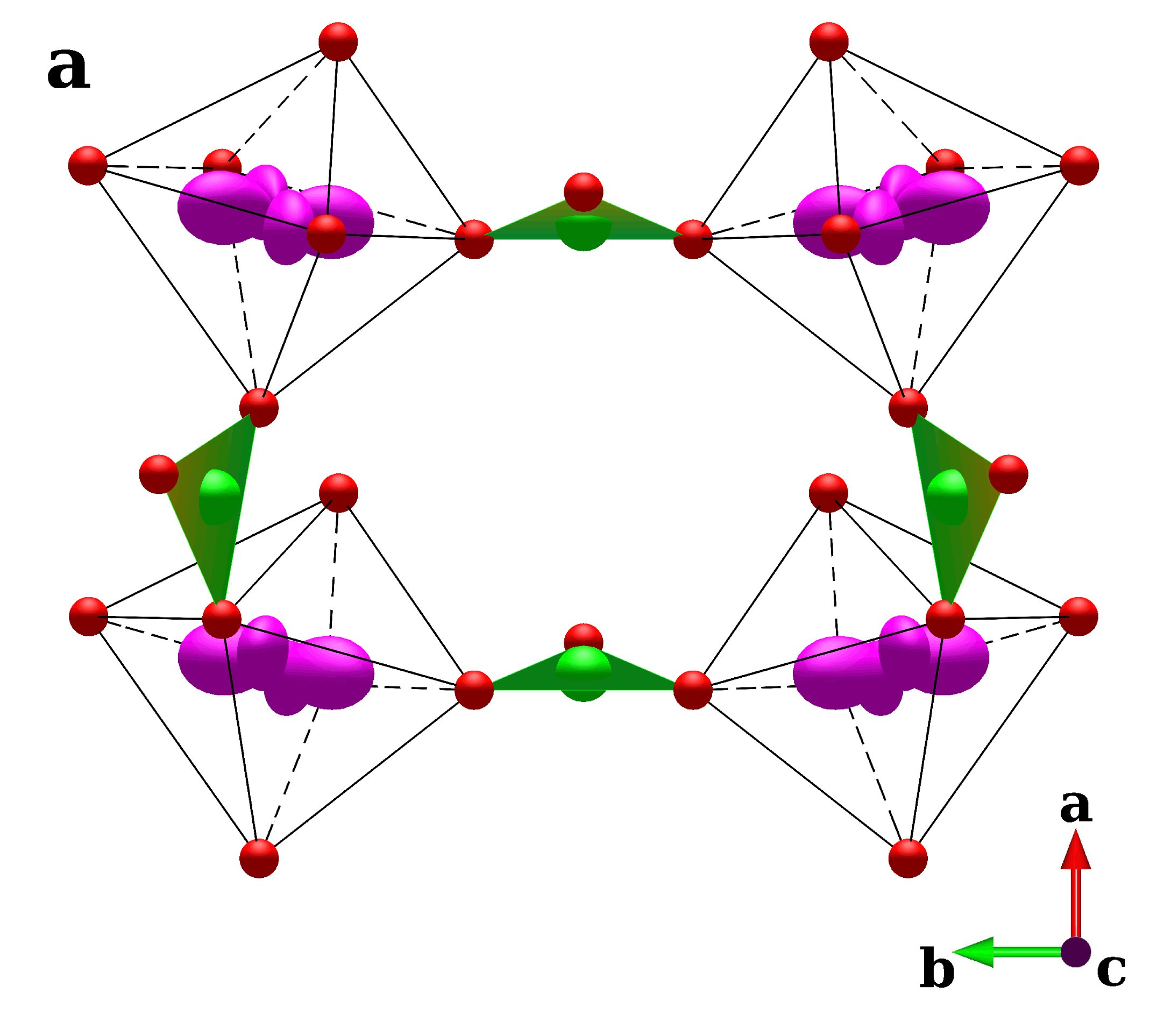}}
\resizebox{7cm}{!}{\includegraphics{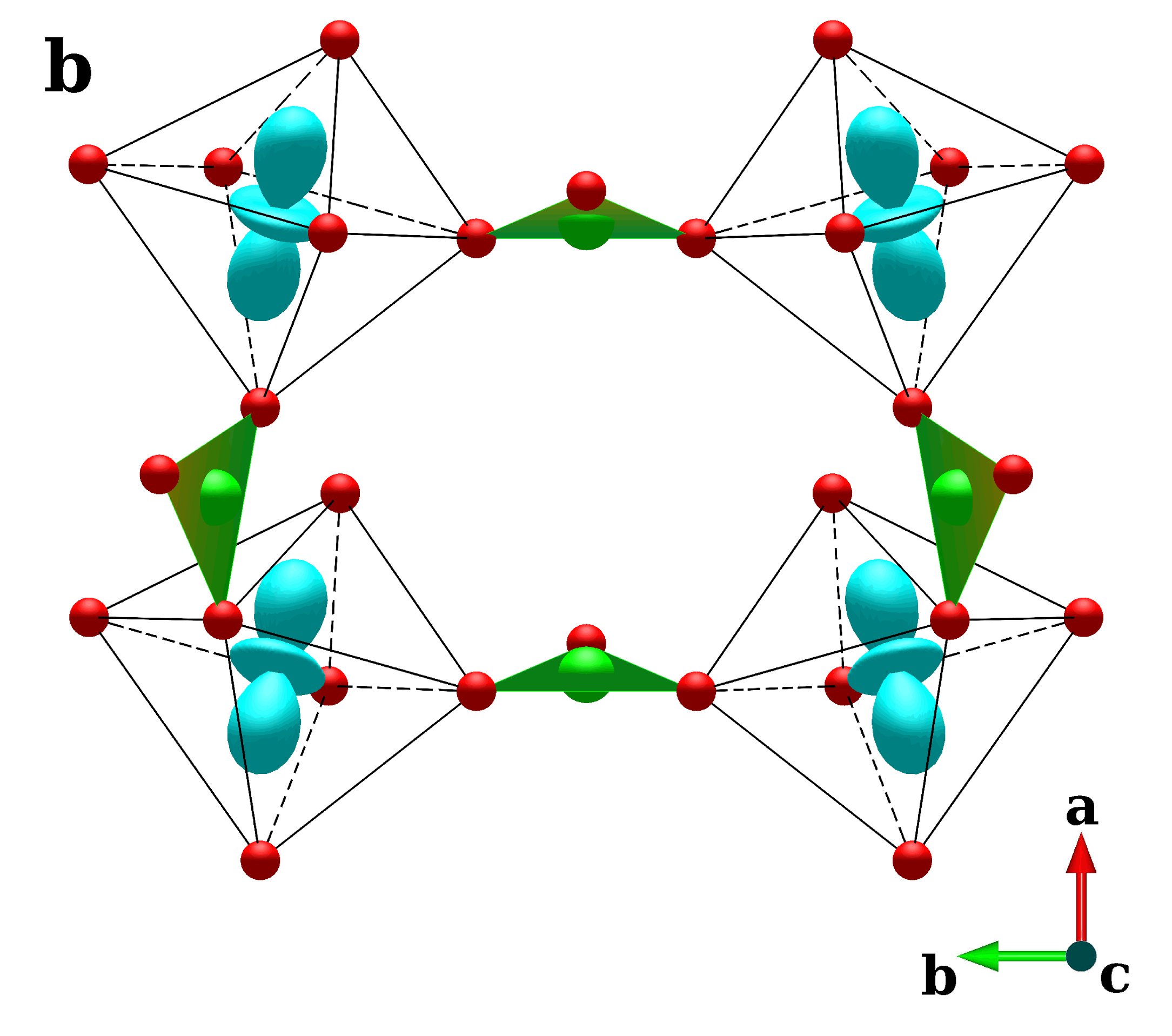}}
\caption{\label{orbit} The $e_g$ orbitals corresponding to the largest elements in the matrix of exchange integrals for exchange along the rung (a) and leg (b) of the ladder determine $J_1$ and $J_2$, respectively.}
\end{figure}

In order to understand the quantitative difference between $J_1$ and $J_2$, we are going to analyze the partial orbital contributions from different Ni-$d$ orbitals to the matrix of the exchange integrals. The largest elements in this matrix for the $J_1$ and $J_2$ pathways correspond to the exchange between the orbitals shown in Fig.~\ref{orbit}a and b, respectively. From Fig.~\ref{orbit}a, it is clearly seen that these orbitals have $x^2$-$y^2$-like symmetry and their lobes are pointing towards the oxygen atoms along $b$ axis forming $\sigma$-bonds. The $p$-orbitals of the oxygen ions are strongly overlapping with these orbitals which leads to a strong antiferromagnetic interaction $J_1$.

The orbitals which determine the $J_2$ exchange along the legs of the ladder have  3$z^2$-$r^2$-like symmetry (Fig.~\ref{orbit}b). $J_2$ is much smaller than $J_1$ since the lobes of the 3$z^2$-$r^2$-like orbitals on the two Ni$^{2+}$-ions along the legs (i.e., in {\it a}-direction) tilt in different directions. The $p$-orbital pointing towards the 3$z^2$-$r^2$-orbital on one particular Ni$^{2+}$-ion will be directed almost perpendicular to the 3$z^2$-$r^2$-orbital of the neighboring Ni$^{2+}$-ion in the leg. Such geometry results in a rather weak 3$z^2$-$r^2$-O $p$ overlap leading to an exchange constant seven times smaller than $J_1$. In comparison, the inter-ladder interactions $J_{3}$ and $J_{4}$ are negligible since there are no nitrate groups between Ni atoms in these directions and the Ni-Ni distances are large, i.e., 7.3 and 6.8~\AA, respectively.

These TB-LMTO-ASA results concerning the isotropic exchange interactions are confirmed by the GGA+U VASP calculations. We have evaluated the exchange integral along the rungs ($J_1$) by the total energy difference of ferromagnetic and antiferromagnetic configurations, $J_{1}$=$\frac{E_{FM}-E_{AFM}}{8S^2}$. The antiferromagnetic configuration has the lower total energy and the magnetic moment on the Ni$^{2+}$ ion is 1.78~$\mu_B$ in accordance with the results of the LMTO calculations. The obtained energy gap, however, amounts to 3.1 eV and is larger than the one obtained within LMTO. The value of the exchange interaction $J_1$=12.73 K agrees with the TB-LMTO-ASA results. 

The center of inversion in the \emph{Pnma} structure of Rb$_3$Ni$_2$(NO$_3$)$_7$ is placed at the origin. Since two Ni atoms belonging to the rungs or the legs of the ladder do not transform into each other under inversion operation, one could expect a non-zero Dzyaloshinskii-Moriya interaction for these two pathways. Indeed we obtain $\vec D_1$=(0.01, 0, 0.01)~K and $\vec D_2$=(-0.01, 0.1, 0.08)~K. Note, however, that our experimental data on the powder sample do not allow evidencing such a small effect of the DM exchange.

Since almost zero anisotropy was obtained within the perturbation theory in the framework of the LMTO calculation we try to estimate its value using the VASP package. The single ion anisotropy could be evaluated as total energies difference of several magnetic configurations. Lets us consider the spin on the first Ni site. If the spin has an easy-axis with local $z'$ axis the corresponding term in the Hamiltonian (\ref{Heisenberg}) has the form $A_1S^2_{z'}$. Calculating the total energies of four spin states in which the spin on the first Ni ion directs along $z'$, -$z'$, $x'$, -$x'$ while the spins on other seven Ni ions are along $y'$ direction, one can evaluate the single ion anisotropy parameter as $A_1$=$\frac{E_1+E_2-E_3-E_4}{2S^2}$~\cite{Xiang2011}. In the case of Rb$_3$Ni$_2$(NO$_3$)$_7$ the direction of the easy-axis is unknown. Hence we perform the total energy calculations for four different configurations: in (1) and (2) the spin on the first Ni site is set along the $x$ and $y$ directions, respectively, while all the rest spins are along $z$; in (3) and (4) the spin on the first Ni site is set along $x$ and $z$ directions, respectively, while all the other spins are directed along $y$. The obtained total energies are $E_1$=-783.06992455 eV, $E_2$=-783.06970414 eV, $E_3$=-783.06990379 eV, $E_4$=-783.06987355 eV. One can conclude that the $x$ direction is preferable and a rough estimation of anisotropy could be obtained as $E_1$-$E_2$=-2.6~K. This value still underestimates the experimental result of A=-8.6~K. The estimation of the single ion anisotropy value from electronic structure calculations is burdened by the necessity to evaluate small differences of large total energies. Since there are 132 atoms in the unit cell of Rb$_3$Ni$_2$(NO$_3$)$_7$ it is hard to test the obtained results with respect to parameters responsible for the accuracy of the calculation. 

\section{Comparison with the experiment \label{comp}}
To compare theoretical results with experiment we employ the expression for the Curie-Weiss temperature $\Theta$=$J_0 S(S+1)/3k_B$ obtained by using the high-temperature expansion of the magnetic susceptibility. Here, $J_0$ is the summarized exchange interaction of a given Ni site with the magnetic environment, which is $J_0=J_1+2J_2+2J_3+2J_4$ in case of Rb$_3$Ni$_2$(NO$_3$)$_7$. The value of the Curie-Weiss temperature recalculated from the numerically obtained exchange constants equals -9.6~K (for $U$=6.5~eV) which is in a good agreement with the experimental data of $\Theta = -10.9$~K.

\begin{figure}[!htb]
\resizebox{8cm}{!}{\includegraphics[angle=-90]{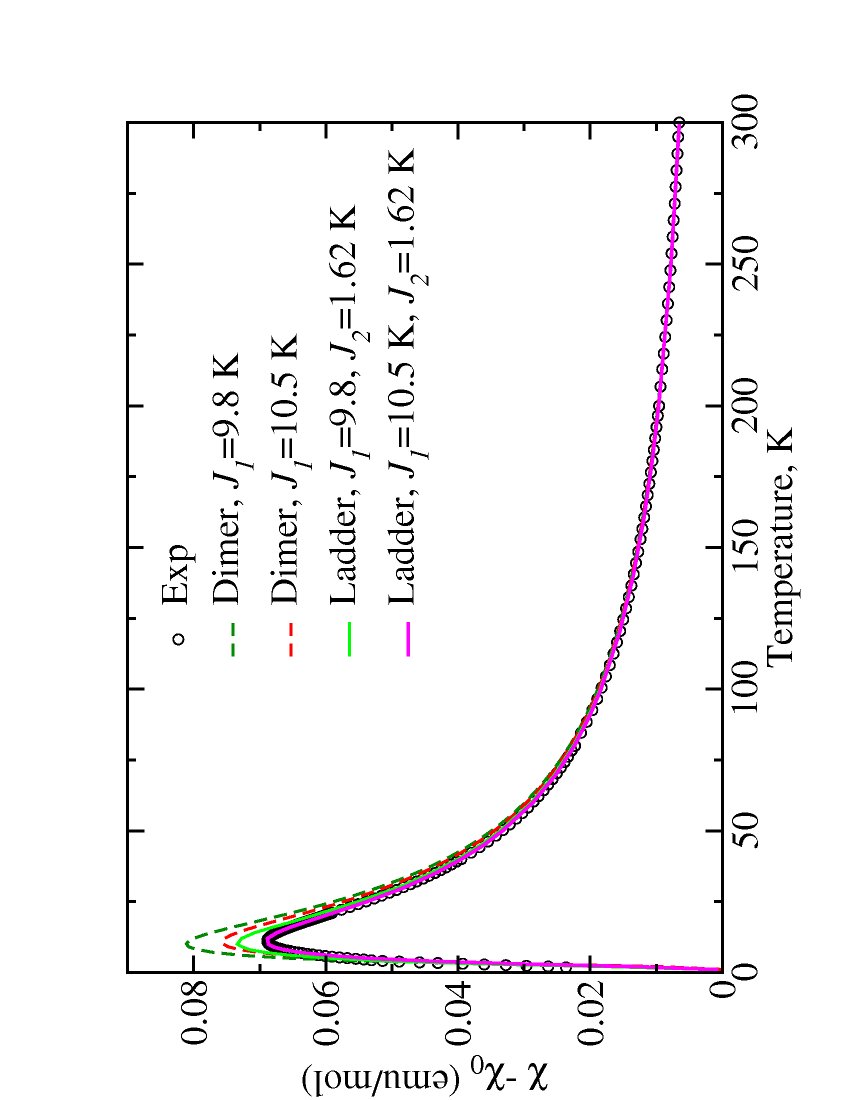}}
\caption{\label{chi_model} Magnetic susceptibility obtained by solving the Heisenberg model for dimers with $J$= 9.8 K and $J$=10.5 K (dark green and red dotted curves, respectively) and for the two-leg ladder model with $J_1$=9.8 K, $J_2$=1.62 K and $J_1$=10.5 K, $J_2$=1.62 K (green and pink solid curves, respectively). Experimental data are shown as open circles.}
\end{figure}

To reproduce the experimentally measured magnetic susceptibility we have solved the Heisenberg model for the $S=1$ two-leg ladder by using the quantum Monte Carlo loop algorithm~\cite{Alet2005} implemented in the ALPS simulation package~\cite{Bauer2011, Albuquerque2007}. The obtained susceptibility for two sets of exchange parameters $J_1$=9.8~K, $J_2$=1.62~K and $J_1$=10.5~K, $J_2$=1.62~K is presented in Fig.~\ref{chi_model} (solid curves). Since the ESR data are described reasonably well within the dimer model we also calculate the susceptibility of independent dimers with the intradimer exchange interactions $J$=9.8~K and $J$=10.5~K by means of full diagonalization (dotted curves). The best agreement with the experimental data is achieved for the ladder model with $J_1$=10.5~K and $J_2$=1.62~K.

\section{Conclusion}
In this paper we report the synthesis, crystal structure and thermodynamic properties of the new Rb$_3$Ni$_2$(NO$_3$)$_7$ compound. The magnetic excitations have been examined in high field high frequency ESR. The measurements of magnetization, specific heat and high-field electron spin resonance have revealed a spin-liquid behavior. From analysis of experimental magnetic susceptibility data and electronic structure calculations the main exchange interaction along the rung of the ladder is estimated to be 9.8-11~K, while the interaction along the leg is about seven times smaller. 
Hence, we can identify these compound as the physical realization of the strong-rung spin-1 ladder model. Even the simple model of the free dimers describes some of the experimental observations, such as the main features of the HF-ESR and the magnetization at low temperature. However, the susceptibility data is better reproduced within ladder model. Hence we suggest that probably a more sophisticated model is needed to obtain the complete agreement of theory and experiment. Such detailed information on thermodynamic properties and values of exchange interactions as well as strong single ion anisotropy would be useful in verifying results of model investigations of spin-1 two-leg ladders.

\section*{Acknowledgements}
We thank A. A. Tsirlin for useful discussions. The work of ZVP and VVM was supported by the Russian Scientific Foundation through the project 14-12-00306. JW Acknowledges support by the IMPRS-QD. ANV and OSV acknowledges support by Russian Foundation for Basic Research Grants 16-02-00021 and 17-02-00211.This work was supported in part from the Ministry of Education and Science of the Russian Federation in the framework of Increase Competitiveness Program of NUST ``MISiS'' (grants K4-2015-020 and K2-2016-066) and by Act 211 of the Government of Russian Federation, agreement 02.A03.21.0006.

\end{document}